# Emergence of charge order from the vortex state of a high temperature superconductor

Tao Wu[1], Hadrien Mayaffre[1], Steffen Krämer[1], Mladen Horvatić[1], Claude Berthier[1], Philip. L. Kuhns[2], Arneil P. Reyes[2], Ruixing Liang[3,4], W.N. Hardy[3,4], D.A. Bonn[3,4], and Marc-Henri Julien[1]

1 Laboratoire National des Champs Magnétiques Intenses, UPR 3228, CNRS-UJF-UPS-INSA, 38042 Grenoble, France

2 National High Magnetic Field Laboratory, Department of Physics, Florida State University, Tallahassee, Florida 32310, USA

3 Department of Physics and Astronomy, University of British Columbia, Vancouver, BC V6T 1Z1, Canada

4 Canadian Institute for Advanced Research, Toronto, Ontario M5G 1Z8, Canada

**Evidence is mounting that charge order competes with superconductivity in high $T_c$ cuprates. Whether this has any relationship to the pairing mechanism is unknown since neither the universality of the competition nor its microscopic nature has been established. Here using nuclear magnetic resonance, we show that, similar to La-214, charge order in $YBa_2Cu_3O_y$ has maximum strength inside the superconducting dome, at doping levels $p$=0.11–0.12. We further show that the overlap of halos of incipient charge order around vortex cores, similar to those visualised in Bi-2212, can explain the threshold magnetic field at which long-range charge order emerges. These results reveal universal features of a competition in which charge order and superconductivity appear** **as joint instabilities of the same normal state, whose relative balance can be field-tuned in the vortex state.**

## INTRODUCTION

The enhancement of spin-stripe order by a magnetic field in La-214 (ref. 1) and the enhanced modulation of the local density of states (LDOS) around the vortex cores in Bi-2212 (ref. 2) are milestone results which have promoted the idea of electronic ordering that competes with superconductivity[3-7]. Recently, the discovery of charge order in $YBa_2Cu_3O_y$ appearing only for fields perpendicular to the copper-oxide planes sufficiently strong to be detrimental to superconductivity[8], as well as the subsequent observation of related charge density wave (CDW) fluctuations[9-11], have provided further evidence of competition. The unambiguous observation of charge order, without spin order, in this compound with a low level of disorder is significant as it reveals the ubiquity of charge ordering tendencies in the normal state of cuprate superconductors. Furthermore, it gives unprecedented opportunity to follow, specifically by NMR, how charge order arises and develops starting deep in the superconducting state where transport techniques are inoperative, through the vortex melting field $H_{melt}$ where ultrasound measurements are blurred, up to field values comparable to $H_{c2}$ which have so far been out of reach for scattering and tunnelling techniques. Here we show with NMR that charge order, *i.e.* a static, long-range, spatial modulation of the charge density, emerges above a threshold magnetic field in the vortex-solid state and we show how this result can be related to earlier evidence of competing orders, thereby highlighting universal aspects of the competition between superconducting and charge orders in cuprates. These results constrain theories relating cuprate superconductivity to the charge instability.

## RESULTS

**Field-dependence of charge order**

Charge order in YBa$_2$Cu$_3$O$_y$ modifies the NMR lineshapes of some of the copper and oxygen ($^{63}$Cu and $^{17}$O) sites in CuO$_2$ planes (ref. 8 and Fig. 1). Here, we adopt the simplest description of these modifications, namely a line splitting[8]. Should the actual lineshape in the charge ordered state be more complex than a simple splitting, this would affect the discussion of the exact pattern of charge order, but not the conclusions of this article which are independent of such details. The spectral modifications induced by the charge order are relatively small, so that detecting a possible departure from a splitting or hypothetical field-dependent modifications (for instance due to a field-dependent ordering wave-vector) is currently beyond our experimental resolution.

The net line splitting involves a splitting $\Delta\nu_{\text{magn}}$ of magnetic hyperfine origin and a splitting $\Delta\nu_{\text{quad}}$ of electric quadrupole origin. Since $\nu_{\text{quad}}$ at Cu and O sites in the cuprates is a linear function of $p$ (ref. 12), we take $\Delta\nu_{\text{quad}}$ to be, in the first approximation, a measure of a charge density difference, *i.e.* the amplitude of the charge order. The expression 'charge order' will be used here generically, with no regard given to its specific morphology, *e.g.* uni- or bi-directional, or its microscopic origin such as Fermi surface instability, electron-phonon coupling or strong correlation effect.

The sharp temperature dependence of the NMR splitting below the onset temperature $T_{\text{charge}}$ (Fig. 2a and ref. 6) and recent ultrasound measurements [13] already indicate that charge order occurs through a phase transition. However, up to now, the field dependence for $T \ll T_{\text{charge}}$ has not been accessed. The new central result is our observation at $T \approx 3$ K of a sharp square-root type increase of $\Delta\nu_{\text{quad}}$, $\Delta\nu_{\text{quad}} \propto (H-H_{\text{charge}})^{1/2}$, starting above a threshold field $H_{\text{charge}} \approx 10.4$ T in a sample having $p=0.109$ and ortho-II oxygen order (Fig. 3a). Qualitatively similar results, albeit covering a smaller field range, were obtained for three other samples from either $^{63}$Cu or $^{17}$O NMR (Fig. 3b-d). This constitutes the first example of a (apparently second-order) quantum phase transition, controlled by the magnetic field, from a homogeneous *d*-wave superconductor to a superconductor with charge-order. Remarkably, the low values of the magnetic hyperfine shift $K$ (Fig. 2c) reveal that the spin-singlet correlations characteristic of the pseudogap state retain their strength in high fields. Charge ordering thus leaves the pseudogap intact.

**Relationship with other probes of charge order in $YBa_2Cu_3O_y$**

The finite value of $H_{charge}$ suggests that there is no static long range charge order in zero field, in agreement with the interpretation of X-ray results in terms of charge-density-wave fluctuations[9]. We note that our $H_{charge} \approx 9.3$ T for $p = 0.12$ corresponds approximately to the field above which the intensity and the width of the superlattice peaks in X-ray measurements[10] become larger in the low temperature limit ($T = 2$ K) than at 66 K, that is at the zero-field $T_c$. This suggests that $H_{charge}$ corresponds to a threshold in the screening of the CDW correlations by the superconducting regions of the sample (see next section for a more precise interpretation of this threshold). On the other hand, sound velocity data[13] for $p = 0.108$ suggest $H_{charge} \approx 18$ T ($c_{11}$ mode) and $H_{charge} \approx 16\pm2$ T (other modes), both larger than $H_{charge} \approx 10.4\pm1.0$ T in NMR for $p=0.109$. It is possible that NMR somewhat underestimates $H_{charge}$ if pre-transitional effects modify the lineshape below the real $H_{charge}$. This should lead to some caution regarding the precise value of $H_{charge}$ but it does not affect the conclusions of this paper.

**Relationship with vortex physics**

For all samples, $H_{charge} \approx 9$-15 T is found to be lower than the melting transition of the vortex lattice which takes place at $H_{melt} > 20$ T for $T \leq 5$ K (ref. 14). The charge-ordering transition thus occurs *inside* the vortex-solid phase. Since the vortex cores represent normal regions of radius $\xi_{SC}$ within the superconductor, it is expected [15-17] that the charge fluctuations detected above $T_c$ (refs. 9-11) continue to develop at low temperatures within the cores where they escape the competition with superconductivity. As suggested by LDOS modulations in Bi-2212 (ref. 2), halos of incipient charge order are centred on the cores and they extend over a typical distance $\xi_{charge} > \xi_{SC}$ (Fig. 4). In these halos, the order is presumably fluctuating faster than the NMR splitting frequency $\Delta\nu$ or is at most weakly pinned so that any static modulation remains weak. With increasing the field, the long-range, static, charge order is expected to appear when these halos start to overlap. This should occur at $H_{charge} = \Phi_0/(2\pi\xi_{charge}^2)$ since the halo density equals the density of vortices whose cores start to overlap at the upper critical field $H_{c2} = \Phi_0/(2\pi\xi_{SC}^2)$. Owing to our observation of a field-induced transition to the charge-ordered state, this prediction is now confirmed by experiments for the first time: $H_{charge} = 9.3\pm1.3$ T for $p = 0.12$ (ortho-VIII) translates into $\xi_{charge} = 16a$, where $a$ is the planar Cu-Cu distance. This is to be compared to $\xi_{charge} \approx 19a$ measured at $H = 9$ T and $T = 2$ K by X-ray diffraction for the same doping level[10]. Despite the obvious simplistic nature of the description (for instance, neither a coupling between $CuO_2$ planes nor an in-plane anisotropy of $\xi_{charge}$ are considered), this agreement is a strong support of its validity. This is the second central result of this work.

**Doping dependence of charge order around p=0.11–0.12.**

On increasing the field further in the $p = 0.109$ sample, $\Delta\nu_{quad}$ saturates at fields of 30-35 T (Fig. 3a). Remarkably, this field scale is similar to $H_{c2}$, as defined from transport measurements in these samples[14,18-20], indicating that the growth of the amplitude of the charge order is controlled by the decrease of the superconducting order parameter. While the precise value of $H_{c2}$ is a matter of debate[21], a dip of $H_{c2}$ near $p = 0.115$-$0.12$ and values in the range 24-60 T appear to be robust conclusions. Furthermore, this dip of $H_{c2}$ is paralleled by both a minimum of $H_{charge}$ (Fig. 3e) and by a maximum of $T_{charge}$ near $p = 0.115$-$0.12$ (Fig. 2b). These correlations are again a manifestation of the competition between superconductivity and charge order and they are consistent with charge order being strongest near $p = 0.115$-$0.12$ (ref. 22) which is within the superconducting dome.

This last observation is, however, puzzling. On one hand, there is an obvious parallel with the stabilisation of charge stripes of period $\lambda \approx 4a$ at a similar (but not strictly identical) doping of $p = 0.125 = 1/8$ in La-214 and possibly in Bi-2212 (ref. 23). This similarity would be natural if the charge density wave in $YBa_2Cu_3O_y$ ($p \approx 0.12$) also had $\lambda \approx 4a$. On the other hand, if the charge order is incommensurate with $\lambda \approx 3.3a$ (refs. 9-11), there is no obvious reason why it should be more stable near $p \approx 0.12$, especially since the chain-oxygen order does not seem to play a prominent role in determining $H_{charge}$ given our finding of similar $H_{charge}$ values in ortho-II ($p = 0.109$) and ortho-VIII ($p = 0.12$) samples (Fig. 3e).

**DISCUSSION**

Even if our understanding of the charge order in $YBa_2Cu_3O_y$ (and perhaps of the 1/8 problem in general) is incomplete, the results reported here reveal an outstanding universality of magnetic field effects in underdoped cuprates. Indeed, a competition between superconductivity and charge order has also been argued[6,15-17] to provide a natural explanation of the field-enhanced spin and charge orders in La-214 (refs. 1,3-6) and LDOS modulations in Bi-2212 (refs. 2,7). Despite possible differences in the morphology of the charge order that may depend on the crystallographic structures and on the level of disorder in different families of cuprates, the above experiments and the results in $YBa_2Cu_3O_y$ are all consistent with the idea that a magnetic field applied perpendicular to the $CuO_2$ planes generates vortices around which fluctuating or weakly pinned, short range charge order is revitalised. The long-range order that should follow from the charge instability of the normal state is initially hindered by superconductivity but it eventually sets in when a sufficiently high density of vortices is reached. On the other hand, when charge order and superconductivity already coexist in zero field, the charge order is simply enhanced by the field[6].

This field-tuned competition differs from the simple coexistence of charge-density-wave order and superconductivity in *e.g.* NbSe$_2$ for which the onset of superconductivity occurs well below the CDW transition and does not affect CDW order. Therefore, the magnetic field has no effect on the CDW[24]. The competition in cuprates is apparently also different from the competition and coexistence of two adjacent phases in the phase diagram of other unconventional superconductors: in this case, the transition to the competing phase can take place at much higher temperature than superconductivity which occurs near the verge of this competing phase[25]. Here, in contrast, the maximum of $T_{charge}$ (and $H_{charge}$) occurs at $p = 0.11$-$0.12$ within the superconducting dome and the transition temperatures are similar ($T_c \sim T_{charge}$), indicating that the two orders have very close energy scales near $p = 0.11$-$0.12$. Such near-degeneracy suggests that, although competing, charge order and superconductivity are joint instabilities of the same normal (pseudogap) state in this doping range.

Despite the mounting evidence for a charge ordering instability in virtually all cuprate families[2,6,8-11,23,26-30], there is still a long way to go before elucidating the importance of this instability in determining the properties of the cuprates. Neither the possibility of an intertwining of the charge and superconducting order parameters[31] nor the possibility of a direct relationship between the two orders[32,33] can be addressed by the present results. Furthermore, more work is needed to understand whether the competition between charge ordering and superconductivity is fundamentally, or only superficially, different from that in other systems showing CDW and superconducting orders in their phase diagrams[34,35]. The most pressing question for clarifying these issues is now to determine how far the charge instability extends in the temperature *vs.* doping phase diagram.

**METHODS**

**Samples**

High quality, oxygen-ordered (Supplementary figure S1), detwinned single crystals of YBa$_2$Cu$_3$O$_y$ were grown in non-reactive BaZrO$_3$ crucibles from high-purity starting materials [36]. Two samples were enriched with the oxygen-17 ($^{17}$O) isotope which, unlike $^{16}$O, possesses a nuclear spin. Table 1 summarises the properties of the four samples studied in this work. Note that the *p* values are obtained on the basis of the values of the superconducting transition temperature $T_c$, as measured by SQUID, using the well-known calibration[37]. Since this calibration was established for samples with $^{16}$O, the isotope effect on the transition temperature $T_c$ of the $^{17}$O-enriched samples must be taken into account. Since $^{16}$O-$^{18}$O exchange produces a change $\Delta T_c \approx -2$ K in the Y-123 system with $T_c$ values around 60 K (ref. 38), the hole content of the two $^{17}$O-enriched crystals was estimated using a $T_c$ value corrected by $\Delta T_c = +1$ K with respect to their $T_c$ value measured by SQUID. This is based on the standard expression for the isotope effect $\Delta T_c/T_c = -\alpha \Delta M/M$ where $M$ is the isotope mass, and $\alpha$ is the isotope effect exponent ($\alpha \approx 0.27$ here in YBa$_2$Cu$_3$O$_y$). The correction for $^{16}$O→$^{17}$O exchange is thus half of that for $^{16}$O→$^{17}$O.

**NMR spectra of quadrupolar nuclei**

The resonance frequency of a given ($m \leftrightarrow m$-1) transition for a nucleus of spin $I > 1/2$ is given by the sum of magnetic hyperfine and electric quadrupole contributions[39]:

$$\nu(m \leftrightarrow m-1) = \nu_{\text{magn}} + \nu_{\text{quad}}(m \leftrightarrow m-1), \qquad (1)$$

with $-I+1 \leq m \leq I$, $\nu_{\text{magn}} = (1 + K_{zz})\,\nu_{\text{ref}}$, where $K_{zz}$ is the component of the hyperfine magnetic shift tensor $\mathcal{K}$ along the magnetic field direction and $\nu_{\text{ref}} = \gamma H$ is a reference frequency (such as the resonance frequency of the bare nucleus in vacuum or the resonance frequency of the nucleus in a substance without unpaired electrons).

To first order in perturbation (although not negligible, the second order is not necessary for the simple qualitative background that we intend to provide here),

$$\nu_{\text{quad}} = \tfrac{1}{2}\left(m - \tfrac{1}{2}\right)(3\cos^2\theta - 1 - \eta\sin^2\theta\,\cos 2\phi)\nu_Q, \qquad (2)$$

where $\nu_Q = \frac{3e^2qQ}{2I(I-1)h}$ is the quadrupole frequency, and $eq = \frac{\partial^2 V}{\partial^2 z} = \mathcal{V}_{zz}$ where $V$ is the electrostatic potential at the nucleus position and $\mathcal{V}$ is the corresponding electric field gradient tensor. The principal axes (X,Y,Z) of the $\mathcal{V}$ tensor are defined such that $|\mathcal{V}_{XX}| \leq |\mathcal{V}_{YY}| \leq |\mathcal{V}_{ZZ}|$ and the asymmetry parameter $\eta = (\mathcal{V}_{XX} - \mathcal{V}_{YY})/\mathcal{V}_{ZZ}$. $\theta$ is the polar angle between the magnetic field direction $z$ and $\mathcal{V}_{ZZ}$ and $\phi$ is the azimuthal angle in the $(x,y)$ plane perpendicular to the field. For planar $^{63}$Cu, Z is the crystalline *c*-axis, while for planar $^{17}$O Z is the Cu-O-Cu direction (*i.e. a* for O(2) and *b* for O(3)). Supplementary figure S2 shows a sketch of a typical quadrupole-split NMR spectrum.

Understanding precisely how the charge density modulation affects the electric field gradient and thus modifies $\Delta\nu_{\text{quad}} = f(\nu_Q, \eta, \theta, \phi)$ requires complex ab-initio calculations which are beyond the scope of this article. Since $\nu_{\text{quad}}$ at Cu and O sites in cuprates is a linear function of $p$ (ref. 13), we take $\Delta\nu_{\text{quad}}$ to be, in the first approximation, a measure of a charge density difference, that is the amplitude of charge order.

**NMR methods**

Experiments were performed in the LNCMI-Grenoble resistive magnets M1, M9 and M10 as well as in the NHFML hybrid magnet. Standard spin-echo techniques were used with heterodyne spectrometers. Spectra were obtained at fixed magnetic fields by combining Fourier transforms of the spin-echo signal recorded for regularly-spaced frequency values[40].

The $^{27}$Al NMR reference signal from metallic aluminium was used to calibrate the external magnetic field values. The $^{17}K$ measurements were performed with $H\|c$, on the central line, that is the $(-1/2 \leftrightarrow 1/2)$ transition, where O(2) and O(3) sites overlap. The $^{17}K$ values thus represent average values for these two sites. Neither the very weak and broad $^{17}$O signal from the chains nor the sharp signal from apical $^{17}$O sites significantly affected the determination of the position of the O(2,3) central line. $^{17}K$ values are given with respect to the resonance frequency of the bare nucleus and they are in excellent agreement with earlier works[41].

In order to determine the $^{17}$O line-splitting, the magnetic field was tilted by $\theta = 16°$ off the *c*-axis in order to separate O(2) from O(3) satellite transitions. In that case, the quoted magnetic field values correspond to the *c*-axis component (that is they are corrected by a factor $\cos(16°) = 0.961$ with respect to the total external field values), which is justified by the disappearance of charge order when $H \perp c$ (ref. 8).

We report here the field-induced modifications of the O(2) NMR lines, which are those sites from Cu-O-Cu bonds aligned along the *a* axis, which is perpendicular to the chain direction *b*. Clear field-induced spectral modifications are also observed for O(3E) and/or O(3F) sites (planar sites in bonds along *b*, below empty and filled chains, respectively) but these could not be easily analysed since the O(3E) and O(3F) lines overlap. A complete account and interpretation of $^{17}$O NMR spectra in the charge-ordered state is beyond the scope of the present work and will be published separately (Wu, T. *et al*., in preparation).

**Analysis of NMR spectra**

The separation of the magnetic hyperfine, $\Delta\nu_{\text{magn}}$, and quadrupole, $\Delta\nu_{\text{quad}}$, contributions to the total line splitting was performed by reproducing the experimental positions of $^{63}$Cu(2F) or $^{17}$O(2) lines with a simulation based on an exact diagonalisation of the nuclear-spin Hamiltonian. For $^{17}$O(2), it is also possible to extract $\Delta\nu_{\text{quad}}$ by subtracting the total splitting $\Delta\nu_{\text{total}}(1) = \Delta\nu_{\text{magn}} + \Delta\nu_{\text{quad}}$ of the $(1/2 \leftrightarrow 3/2)$ satellite from the total splitting $\Delta\nu_{\text{total}}(2) = \Delta\nu_{\text{magn}} + 2\Delta\nu_{\text{quad}}$ of the $(3/2 \leftrightarrow 5/2)$ satellite. However, $\Delta\nu_{\text{total}}(1)$ is not always experimentally accessible. For $p=0.09$, the saturation of $\Delta\nu_{\text{quad}}$ and $\Delta\nu_{\text{magn}}$ above ~30 T is confirmed by the linear field-dependence of the total splitting $\Delta\nu =$

$\Delta \nu_{\text{quad}} + \Delta \nu_{\text{magn}} = \Delta \nu_{\text{quad}} + \Delta K_{zz}\, \gamma H$, and by the perfect overlap of the lineshapes of the $^{63}$Cu central transition at all fields above 30 T when plotted in a frequency scale normalised by field.

No stable fit of the temperature dependence of $\Delta \nu_{\text{quad}}$ could be performed close to $T_{\text{charge}}$. Nevertheless, we found that the temperature dependence of the guide to the eye shown in Fig. 2a for the $p$=0.125 data also matches very well the data for $p$=0.104 (6) and $p$=0.109. We thus used this guide to determine the transition temperature of the charge-ordered state, $T_{\text{charge}}$.

$H_{\text{charge}}$ is determined by fitting $\Delta \nu_{\text{quad}}$ data to $(H-H_{\text{charge}})^{0.5}$, so that the result does not depend on the points for which $\Delta \nu_{\text{quad}}$ is assumed to be zero. The value of $H_{\text{charge}}$ is largely determined by the curvature of $\Delta \nu_{\text{quad}}$ vs. $H$: that is, the data points for which the large splitting results in relatively small error bars contribute as much as the points close to $H_{\text{charge}}$ which have larger error bars. The data for the $p$=0.109 sample were fit to $(H-H_{\text{charge}})^{\alpha}$ and the fit resulted in $\alpha = 0.49 \pm 0.09$.

Acknowledgements: We thank N. Doiron-Leyraud, J. Hoffman, B. Keimer, S. Kivelson, D. Le Boeuf, M. Le Tacon, C. Proust, L. Taillefer, B. Vignolle, M. Vojta for discussions. Work at Grenoble was supported by the 7th framework programme "Transnational Access" of the European Commission, contract N° 228043 – EuroMagNET II – Integrated Activities, by Pôle SMIng - Université J. Fourier - Grenoble and by the French Agence Nationale de la Recherche (ANR) under reference AF-12-BS04-0012-01 (Superfield). Work at Tallahassee was supported by National Science Foundation Cooperative Agreement No. DMR-0654118, the State of Florida, and the U.S. Department of Energy. Work at Vancouver was supported by the Canadian Institute for Advanced Research and the Natural Science and Engineering Research Council.

Authors' Contributions: R.L., W.N.H. and D.A.B. prepared the samples. T.W., H.M, S.K. P.L.K. A.P.R. and M.-H.J. performed the experiments. S.K., M.H. and H.M. developed the NMR setups in Grenoble. P.L.K and A.P.R. developed the NMR setups in Tallahassee. T.W. and M.-H.J. analysed the data. C.B. provided conceptual advice. M.H.J. wrote the paper and supervised the project. All authors discussed the results and commented on the manuscript.

The authors declare no competing financial interests. Correspondence and requests for materials should be addressed to M.-H.J. (marc-henri.julien@lncmi.cnrs.fr).

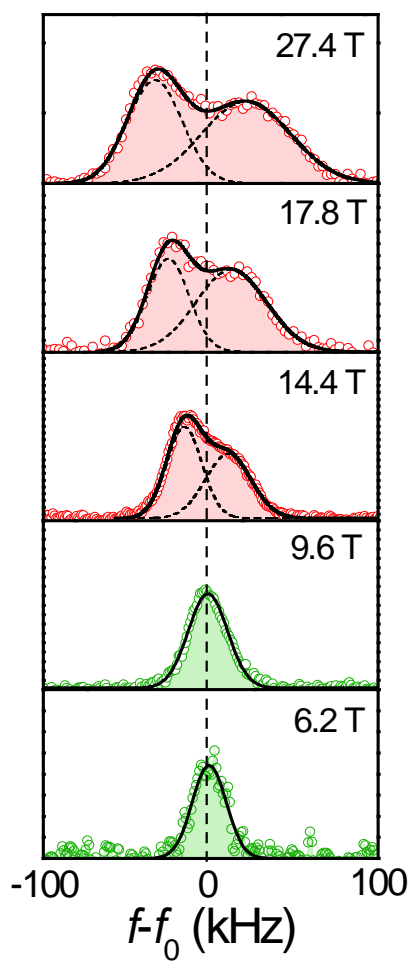

**Figure 1 | $^{17}$O NMR evidence of charge order in YBa$_2$Cu$_3$O$_{6.56}$ ($p$=0.109).**

Magnetic field induced modifications of the highest-frequency quadrupole satellite of O(2) sites (lying in bonds oriented along the $a$ axis) at $T$ = 2.9 K. $f_0$ (~40-160 MHz) is the frequency of the centre of the shown spectrum. Continuous lines are fits with one Gaussian at 6.2 and 9.6 T and with two Gaussian functions (each shown as a dotted line) at higher fields. See Methods section for more information about the $^{17}$O spectra.

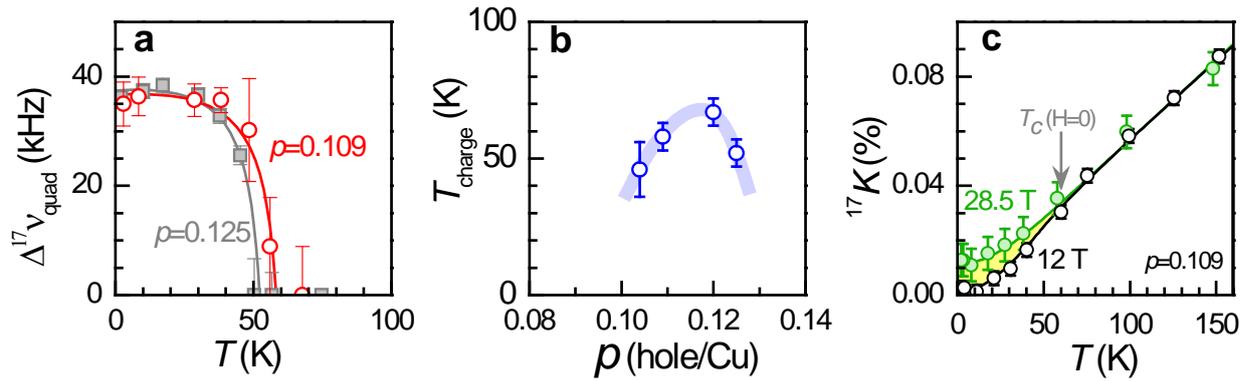

**Figure 2 | Temperature-induced transition towards charge order in the pseudogap state.**

**a**, Quadrupole part of the splitting of the highest-frequency O(2) quadrupole satellite for $p = 0.109$ and $p = 0.125$ (see Methods for details) as a function of temperature in fields of 27.4 and 28.8 T, respectively. The lines are guides to the eye. **b**, Transition temperature $T_{charge}$ showing a maximum around hole doping $p = 0.115$-$0.12$. The thick trace is a guide to the eye. **c**, Magnetic hyperfine shift $^{17}K$ of O(2,3) sites for $p = 0.109$ and $H||c$. No anomalous change of $^{17}K$ is observed across $T_{charge}$. At low temperature in the charge ordered state, the maximum shift variation $\Delta^{17}K \approx 0.01$ % between 12 and 28.5 T (yellow region) represents a minor change compared to the decrease $\Delta^{17}K \approx 0.12$ % between $T = 300$ K and 60 K associated with the pseudogap. The pseudogap is thus essentially unaffected by the occurrence of charge order. This result agrees with the relatively modest size of the field-induced changes in the $^{63}$Cu relaxation rate $1/T_1$ (ref. 8). The field dependence of $^{17}K$ below $T_c$ arises from the density of nodal states in a $d$-wave superconductor[42]. Error bars represent standard deviations of the fit parameters.

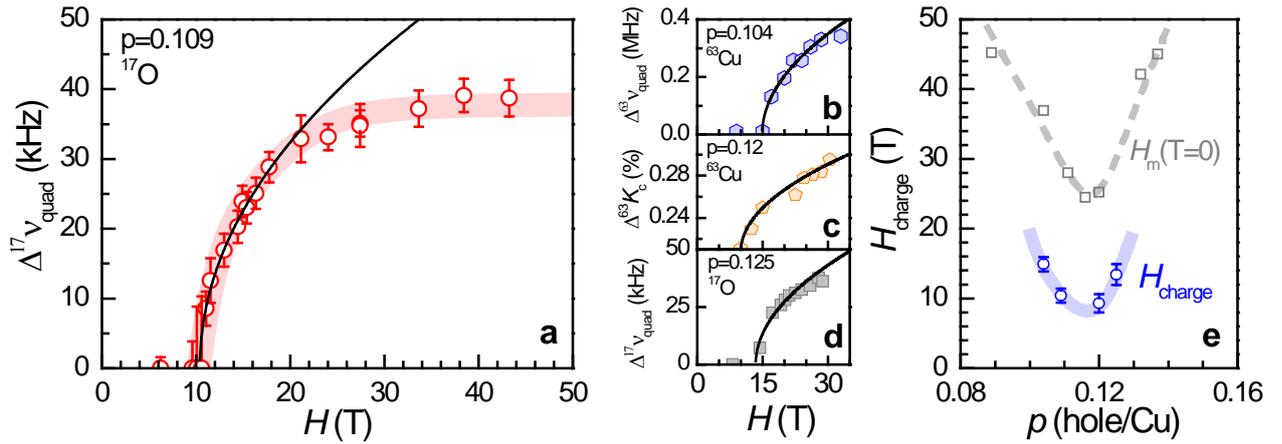

**Figure 3 | Quantum phase transition to the charge ordered state.**

**a**, Quadrupole part of the splitting of the O(2) line shown in Fig. 1 for $p$=0.109 and $T \approx 3$ K (see Methods for details). The thick red trace is a guide to the eye. The thin black line is a fit to $(H-H_c)^{0.5}$. **b**, Quadrupole splitting of $^{63}$Cu(2F) (planar Cu sites below oxygen-filled chains) for $p$=0.104 (ortho-II) at $T$=2 K. **c**, Full linewidth at half maximum of the $^{63}$Cu central line for $p$=0.12 (ortho-VIII) at $T$=1.4 K. Because of the broad and complex Cu spectra in this sample, the splitting of the $^{63}$Cu satellite line could not be observed[8]. However, the width of the central line reflects the modifications of the lineshape due to charge order modulating the hyperfine fields[8]. **d**, Quadrupole part of the splitting of the $^{17}$O(2) line for $p$=0.125 (ortho-VIII) at $T$=2.1 K. **e**, Hole-doping dependence of the onset field $H_{charge}$ as determined from a fit of the data to $(H-H_{charge})^{0.5}$ (thin black lines in **a-d**). The minimum of $H_{charge}$ near $p$=0.115-0.12 parallels that of the vortex-melting field $H_{melt}(T \to 0)$ which has been argued to reflect the upper critical field $H_{c2}(T \to 0)$ (ref. 15). Error bars represent standard deviations in the fit parameters.

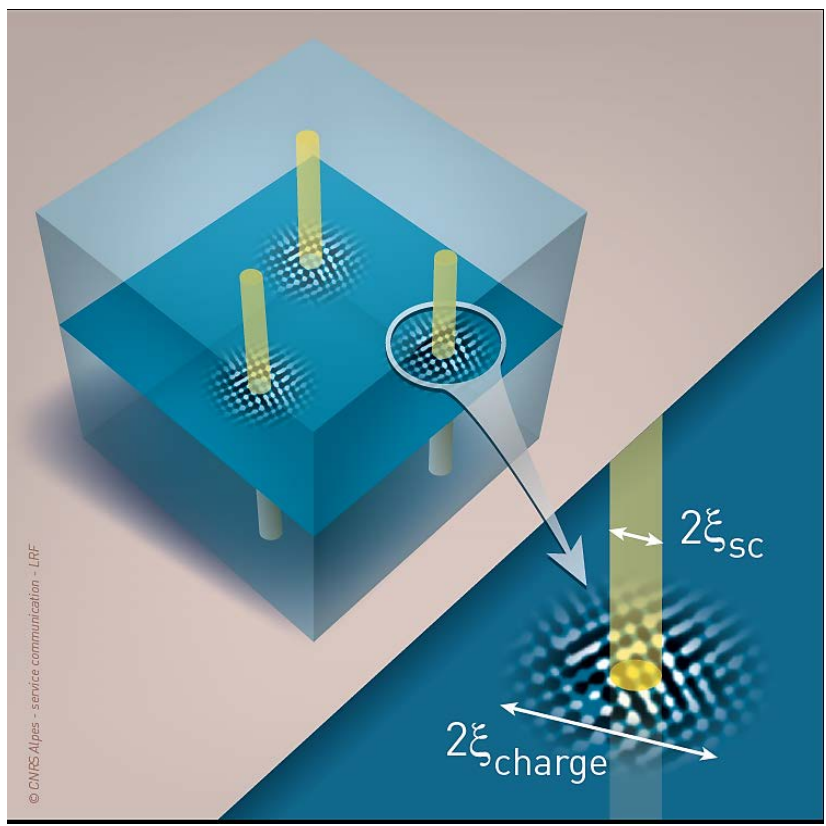

**Figure 4 | Halos of incipient charge order centred on vortex cores.**

The yellow tubes represent the vortex cores of radius given by the superconducting coherence length $\xi_{SC}$. Black and white halos represent a snapshot of sites with high and low charge density. The charge density modulations are presumably fluctuating, or they are at most weakly pinned, for fields lower than $H_{charge}$ and thus do not produce an NMR splitting. These halos of incipient charge order start to overlap at $H_{charge}$ and thus induce long-range order. This description, inspired from scanning tunnelling microscopy results in Bi-2212 (ref. 2), is quantitatively consistent with our results in $YBa_2Cu_3O_y$ (see text). $\xi_{charge}$ is the typical length over which the charge density is correlated, thus defining halos of diameter $2\xi_{charge}$ around the vortex cores. The actual pattern of the charge density modulation plays no role in the discussion of this paper, so its representation is arbitrary here.

| Nominal y in $YBa_2Cu_3O_y$ | Chain-oxygen order | $T_c$ | Doping p (hole/Cu) | Comment |
|---|---|---|---|---|
| **6.54** | O-II | 59.6 K | 0.104 | $T_c$ & p values revised with respect to Ref. (8) |
| **6.56** | O-II | 59.8 K | 0.109 | $^{17}$O enriched |
| **6.67** | O-VIII | 67 K | 0.12 | Ref. (8) |
| **6.68** | O-VIII | 67.8 K | 0.125 | $^{17}$O enriched |

**Table 1 | Samples reported in this study.**